\documentclass[12pt]{article}
\usepackage{axodraw}
\usepackage{epsfig}
\makeatletter
\if@twoside \oddsidemargin -17pt \evensidemargin 00pt
\else \oddsidemargin 00pt \evensidemargin 00pt
\fi
\expandafter\ifx\csname mathrm\endcsname\relax\def\mathrm#1{{\rm #1}}\fi

\renewcommand{\theequation}{\thesection.\arabic{equation}}
\newcounter{saveeqn}

\@addtoreset{equation}{section}

\makeatother

\def\co{\relax}
\def\co{,}
\def\nl{\nonumber\\}
\def\nlc{\co\nonumber\\}

\def\beq{\begin{equation}}
\def\eeq{\end{equation}}
\def\beqar{\begin{eqnarray}}
\def\eeqar{\end{eqnarray}}
\def\barr#1{\begin{array}{#1}}
\def\earr{\end{array}}
\def\bfi{\begin{figure}}
\def\efi{\end{figure}}
\def\btab{\begin{table}}
\def\etab{\end{table}}
\def\bce{\begin{center}}
\def\ece{\end{center}}
\def\nn{\nonumber}

\def\text{\textstyle}

\def\al{\alpha}

\def\ga{\gamma}
\def\de{\delta}

\def\la{\lambda}
\def\si{\sigma}

\def\De{\Delta}

\def\refeq#1{\mbox{(\ref{#1})}}
\def\reffi#1{\mbox{Fig.~\ref{#1}}}
\def\reffis#1{\mbox{Figs.~\ref{#1}}}
\def\refta#1{\mbox{Table~\ref{#1}}}

\def\refse#1{\mbox{Sect.~\ref{#1}}}
\def\refses#1{\mbox{Sects.~\ref{#1}}}

\def\citere#1{\mbox{Ref.~\cite{#1}}}
\def\citeres#1{\mbox{Refs.~\cite{#1}}}

\def\solid{\raise.9mm\hbox{\protect\rule{1.1cm}{.2mm}}}
\def\dash{\raise.9mm\hbox{\protect\rule{2mm}{.2mm}}\hspace*{1mm}}

\newcommand{\GeV}{\unskip\,\mathrm{GeV}}

\newcommand{\TeV}{\unskip\,\mathrm{TeV}}

\def\mathswitchr#1{\relax\ifmmode{\mathrm{#1}}\else$\mathrm{#1}$\fi}

\newcommand{\PW}{\mathswitchr W}
\newcommand{\PZ}{\mathswitchr Z}

\newcommand{\PH}{\mathswitchr H}
\newcommand{\Pe}{\mathswitchr e}

\newcommand{\Pf}{\mathswitch f}

\newcommand{\Pt}{\mathswitchr t}

\newcommand{\Pep}{\mathswitchr {e^+}}
\newcommand{\Pem}{\mathswitchr {e^-}}

\def\mathswitch#1{\relax\ifmmode#1\else$#1$\fi}

\newcommand{\Mf}{\mathswitch {m_\Pf}}

\newcommand{\MW}{\mathswitch {M_\PW}}

\newcommand{\MZ}{\mathswitch {M_\PZ}}
\newcommand{\MH}{\mathswitch {M_\PH}}
\newcommand{\Me}{\mathswitch {m_\Pe}}

\newcommand{\Mt}{\mathswitch {m_\Pt}}

\newcommand{\rw}{\mathrm{w}}
\newcommand{\sw}{\mathswitch {s_\rw}}
\newcommand{\cw}{\mathswitch {c_\rw}}

\newcommand{\GF}{\mathswitch {G_\mu}}

\def\ie{i.e.\ }
\def\eg{e.g.\ }
\def\cf{cf.\ }

\newcommand{\Oa}{\mathswitch{{\cal{O}}(\alpha)}}

\newcommand{\wirk}{\mathrm{\left(\frac{\mathrm{d}\sigma}{\mathrm{d}\Omega}\right)}}

\newcommand{\Brems}{\mathrm{brems}}

\newcommand{\Born}{\mathrm{Born}}
\newcommand{\virt}{\mathrm{virt}}
\newcommand{\Boxen}{\mathrm{B}}
\newcommand{\Ver}{\mathrm{V}}
\newcommand{\elect}{\mathrm{em}}
\newcommand{\w}{\mathrm{w}}
\newcommand{\fer}{\mathrm{fer}}
\newcommand{\bos}{\mathrm{bos}}
\newcommand{\rem}{\mathrm{rem}}
\newcommand{\run}{\mathrm{run}}
\newcommand{\rR}{\mathrm{R}}
\newcommand{\rL}{\mathrm{L}}
\newcommand{\RR}{\mathrm{RR}}
\newcommand{\LL}{\mathrm{LL}}
\newcommand{\LR}{\mathrm{LR}}
\newcommand{\RL}{\mathrm{RL}}
\newcommand{\Sp}{\mathrm{Sp}}

\newcommand{\rd}{{\mathrm{d}}}

\newcommand{\M}{{\cal {M}}}
\renewcommand{\L}{{\cal L}}

\newcommand{\Moeller}{{{M\o ller}}}

\def\Re{\mathop{\mathrm{Re}}\nolimits}

\def\draftdate{\relax}
\def\mda{\relax}
\def\mua{\relax}
\def\mla{\relax}
\def\draft{
\def\thtystars{******************************}
\def\sixtystars{\thtystars\thtystars}
\typeout{}
\typeout{\sixtystars**}
\typeout{* Draft mode!
         For final version remove \protect\draft\space in source file *}
\typeout{\sixtystars**}
\typeout{}
\def\draftdate{\today}
\def\mua{\marginpar[\boldmath\hfil$\uparrow$]%
                   {\boldmath$\uparrow$\hfil}%
                    \typeout{marginpar: $\uparrow$}\ignorespaces}
\def\mda{\marginpar[\boldmath\hfil$\downarrow$]%
                   {\boldmath$\downarrow$\hfil}%
                    \typeout{marginpar: $\downarrow$}\ignorespaces}
\def\mla{\marginpar[\boldmath\hfil$\rightarrow$]%
                   {\boldmath$\leftarrow $\hfil}%
                    \typeout{marginpar: $\leftrightarrow$}\ignorespaces}
\def\Mua{\marginpar[\boldmath\hfil$\Uparrow$]%
                   {\boldmath$\Uparrow$\hfil}%
                    \typeout{marginpar: $\Uparrow$}\ignorespaces}
\def\Mda{\marginpar[\boldmath\hfil$\Downarrow$]%
                   {\boldmath$\Downarrow$\hfil}%
                    \typeout{marginpar: $\Downarrow$}\ignorespaces}
\def\Mla{\marginpar[\boldmath\hfil$\Rightarrow$]%
                   {\boldmath$\Leftarrow $\hfil}%
                    \typeout{marginpar: $\Leftrightarrow$}\ignorespaces}
\overfullrule 5pt
\oddsidemargin -15mm
\marginparwidth 29mm
}

\makeatletter

\def\eqnarray{\stepcounter{equation}\let\@currentlabel=\theequation
\global\@eqnswtrue
\global\@eqcnt\z@\tabskip\@centering\let\\=\@eqncr
$$\halign to \displaywidth\bgroup\hskip\@centering
  $\displaystyle\tabskip\z@{##}$\@eqnsel&\global\@eqcnt\@ne
  \hskip 2\arraycolsep \hfil${##}$\hfil
  &\global\@eqcnt\tw@ \hskip 2\arraycolsep $\displaystyle\tabskip\z@{##}$\hfil
   \tabskip\@centering&\llap{##}\tabskip\z@\cr}
\def\appendix{\par
 \setcounter{section}{0} \setcounter{subsection}{0}
 \def\thesection{\Alph{section}}}

\makeatother

\begin{document}
\thispagestyle{empty}
\def\thefootnote{\fnsymbol{footnote}}
\setcounter{footnote}{1}
\null
\draftdate\hfill  PSI-PR-98-16\\
\strut\hfill hep-ph/9807446
\vskip 0cm
\vfill
\begin{center}
{\Large \bf
Electroweak radiative corrections \\ to polarized \Moeller\ scattering 
at high energies
\par} \vskip 2.5em
{\large
{\sc A.~Denner%
}\\[1ex]
{\normalsize \it Paul Scherrer Institut\\
CH-5232 Villigen PSI, Switzerland}\\[2ex]
{\sc S.~Pozzorini%
}\\[1ex]
{\normalsize \it
Institut f\"ur Theoretische Physik, ETH-H\"onggerberg\\
CH-8093 Z\"urich, Switzerland
}\\[2ex]
}
\par \vskip 1em
\end{center}
\par
\vskip .0cm \vfill {\bf Abstract:} \par 

The cross section for $\Pem\Pem\rightarrow\Pem\Pem$ with arbitrary
electron polarizations is calculated within the Electroweak Standard
Model for energies large compared to the electron mass, including the
complete virtual and soft-photonic $O(\al)$ radiative corrections.
The relevant analytical results are listed, and a numerical evaluation
is presented for the unpolarized and polarized cross sections as well
as for polarization asymmetries. The relative weak corrections are
typically of the order of 10\%. 
At low energies, the bulk of the
corrections is due to the running of the electromagnetic coupling
constant. 
For left-handed electrons, at high energies the vertex and box
corrections involving virtual \PW~bosons become very important. The
polarization asymmetry is considerably reduced by the weak radiative
corrections.
\par
\vskip 1cm
\noindent
July 1998 
\par
\null
\setcounter{page}{0}

\section{Introduction}

The linear electron colliders of the next generation will allow
experiments with highly polarized electron and photon beams starting
from a centre-of-mass energy of a few hundred $\GeV$ up to the $\TeV$
range. The high degree of polarization combined with a large
luminosity provides a powerful tool for suppressing backgrounds and
sorting out interesting physical effects.  The main goal of these
experiments is the search for new phenomena both directly and
indirectly via precision measurements of standard quantities.

The study of \Moeller\ scattering \cite{Mo32},
$\Pem\Pem\rightarrow\Pem\Pem$, offers some particularly interesting
possibilities owing to its large cross section leading to very good
statistics. On the one hand, small angle \Moeller\ scattering can be
used as a luminosity monitor just as Bhabha scattering in $\Pep\Pem$
colliders. On the other hand, the measurement of parity-violating
\Moeller\ asymmetries can be employed for a very precise determination
of the weak mixing angle \cite{Cuypers,Cz98}.  Note that in contrast
to the measurement of the effective mixing angle at the \PZ-boson
resonance, the \Moeller\ asymmetries do not determine the mixing angle
directly.  In higher orders, it can only be extracted from the
measured asymmetries
if the other parameters of the Electroweak Standard Model are kept fixed. 
Based on the lowest-order expressions,
Cuypers and Gambino \cite{Cuypers} have shown that
$\Delta\sin^2{\theta_{\rw}}=\pm0.0001$ may be possible at a $2\TeV$
collider with $90\%$ electron-beam polarization and detector
acceptance down to about $5^\circ$. Simultaneously these measurements
enable a determination of the polarization degree of the electron
beams competitive with Compton polarimetry with a relative precision
below one per cent. This provides an important input for
other precision experiments.

In order to achieve comparable accuracy in experimental measurements
and theoretical predictions, radiative corrections to \Moeller\
scattering have to be taken into account. Within QED, these corrections
have been calculated for the unpolarized cross section in
\citere{QEDcorrunpol} and for polarized electrons in
\citere{QEDcorrpol}. The electromagnetic corrections to \Moeller\
scattering within the Electroweak Standard Model,
\ie including the \PZ-boson-exchange diagrams,
 have been analysed in
\citere{qedmc,emcorr}. The full electroweak corrections have so far only
been calculated at low energies $\sqrt{s}\ll\MZ$ \cite{Cz96}. In this
case,
corrections of $-40\%$ have been found for the polarization asymmetry.

In this paper we present the complete virtual and soft photonic
$\Oa$ electroweak radiative corrections, which have been
obtained by applying crossing symmetry to the results for Bhabha
scattering calculated in \citere{Denner1}.  As in \citere{Denner1},
 we do
not discuss hard-photon corrections and neglect the electron mass
wherever possible. Our results are valid for arbitrary polarizations of the
external electrons.

The paper is organized as follows: In \refse{se:born} we fix our
notation, review the lowest-order predictions and define polarization
asymmetries.  In \refses{se:emcorr} and \ref{se:weakcorr} we give all
analytical results for the electromagnetic and weak radiative
corrections.  
Numerical results for the radiative
corrections are presented and discussed in \refse{se:numeric}. 

\section{Lowest-order predictions}
\label{se:born}

\subsection{Lowest-order cross sections}
We discuss, in the framework of the Electroweak Standard Model
\cite{GSW1}, \Moeller\ scattering at high energies ($\sqrt{s}\gg \Me$). 
The momenta of the incoming and outgoing electrons are denoted by
$p_1,p_2$ and $q_1,q_2$, the corresponding helicities by $\la_1,\la_2$
and $\si_1,\si_2$, respectively. Instead of using
the proper values $\pm1/2$
we indicate the electron helicities simply by $\pm$ or by $\rR,\rL$.

In the centre-of-mass (CM) system the momenta read
\begin{eqnarray}
p_1^\mu&=&(E,0,0,E),\,\qquad q_1^\mu=(E,E\sin{\vartheta},0,E\cos{\vartheta}),\nonumber\\
p_2^\mu&=&(E,0,0,-E),\qquad q_2^\mu=(E,-E\sin{\vartheta},0,-E\cos{\vartheta}),
\end{eqnarray}
where $E$ denotes the energy of the electrons and $\vartheta$ the
scattering angle. The Mandelstam variables are given by
\begin{eqnarray}
s&=&(p_1+p_2)^2=(q_1+q_2)^2=4E^2,\nonumber\\
t&=&(p_1-q_1)^2=(p_2-q_2)^2=-s\frac{1-\cos{\vartheta}}{2},\nonumber\\
u&=&(p_1-q_2)^2=(p_2-q_1)^2=-s\frac{1+\cos{\vartheta}}{2}.
\end{eqnarray}
The interaction of the electrons with the vector bosons
$\gamma,\PZ,\PW$ is described  by left- and right-handed coupling
constants $g^i_\lambda$:
\begin{equation} 
g_\pm^\gamma =1,\qquad g_-^Z =\frac{\displaystyle 2\sw^2-1}{\displaystyle 2\sw\cw},\qquad g_+^Z=\frac{\displaystyle \sw}{\displaystyle \cw},\qquad g_-^W =\frac{1}{\sqrt{2}\,\sw},\qquad g_+^W=0,
\end{equation}
with
\begin{equation}
\cw=\frac{\MW}{\MZ},\qquad \sw=\sqrt{1-\cw^2}.
\end{equation}

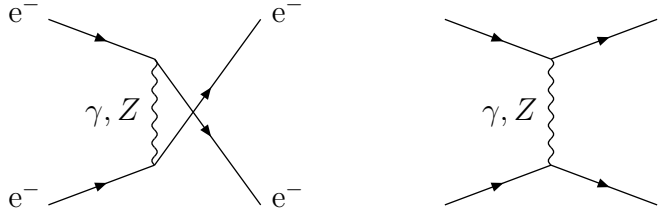
\begin{figure}
\begin{center}
\begin{picture}(250,80)
\put(0,0){
\begin{picture}(100,80)
\Photon(50,20)(50,60){1}{6} 
\ArrowLine(10,5)(50,20)\ArrowLine(50,20)(90,75)
\ArrowLine(10,75)(50,60)\ArrowLine(50,60)(90,5)
\Text(45,40)[r]{$\gamma,Z$}
\Text(-5,9)[l]{$\Pem$}
\Text(-5,79)[l]{$\Pem$}
\Text(107,9)[r]{$\Pem$}
\Text(107,79)[r]{$\Pem$}
\end{picture}}
\put(150,0){
\begin{picture}(100,80)
\Photon(50,20)(50,60){1}{6} 
\ArrowLine(10,5)(50,20)\ArrowLine(50,20)(90,5)
\ArrowLine(10,75)(50,60)\ArrowLine(50,60)(90,75)
\Text(45,40)[r]{$\gamma,Z$}
\end{picture}}
\end{picture}
\end{center}
\caption{\label{borndiagrams}Born diagrams}
\end{figure}

The lowest-order amplitude gets contributions from the exchange of
photons and $\PZ$~bosons in the $u$- and $t$-channel diagrams shown in
\reffi{borndiagrams}. 
The corresponding matrix elements are denoted by
$\mathcal{M}_\Born^{ri}(\lambda_1,\lambda_2,\sigma_1,\sigma_2)$, with
$r=u,t$ and $i=\gamma ,Z$. 
In the ultra-relativistic limit 
$(s, |t|, |u|\gg\Me^2)$, owing to
helicity selection rules, only the following matrix elements are
non-vanishing ($\lambda=\pm$):
\begin{eqnarray}
\mathcal{M}_\Born^{ui}(\lambda,\lambda,\lambda,\lambda)&=\mathcal{M}_1^{ui}(\lambda)=&g^i_{\lambda }g^i_{\lambda }\frac{2s}{u-M_i^2}, \nonumber\\
\mathcal{M}_\Born^{ui}(\lambda,-\lambda, -\lambda,\lambda)&=\mathcal{M}_2^{ui}(\lambda)=&g^i_{+}g^i_{-}\frac{2t}{u-M_i^2},\nonumber\\
\mathcal{M}_\Born^{ti}(\lambda,\lambda, \lambda,\lambda)&=\mathcal{M}_1^{ti}(\lambda)=&g^i_{\lambda }g^i_{\lambda }\frac{2s}{t-M_i^2},\nonumber\\
\mathcal{M}_\Born^{ti}(\lambda,-\lambda, \lambda,-\lambda)&=\mathcal{M}_3^{ti}(\lambda)=&g^i_{+}g^i_{-}\frac{2u}{t-M_i^2}.
\label{bornmatrix}
\end{eqnarray}

Summing and squaring the matrix elements yields the Born cross
sections
in the CM frame 
\begin {equation}
\wirk_{a,\Born}(\lambda)=\frac{\al^2}{4s}
\Bigl|\sum_{r,i}\M_{a,\Born}^{ri}(\lambda)\Bigr|^2; 
\label{borncs0}
\end{equation}
the index $a=1,2,3$ corresponds to the different sets of
helicities in (\ref{bornmatrix}). The cross sections for arbitrary
degrees of longitudinal polarization $P_1,P_2$ of the incoming
electrons and unpolarized outgoing electrons are
 obtained as
\begin{equation}
\wirk(P_1,P_2)=\sum_{\lambda_1,\lambda_2} \frac{1+\lambda_1 P_1}{2}\frac{1+\lambda_2 P_2}{2}\wirk_{\lambda_1,\lambda_2}.
\label{polcs}
\end{equation}
In lowest order, the corresponding polarized cross sections are 
given by
\begin{eqnarray}
\wirk_{\mathrm{\lambda,\lambda}}&=&\wirk_{1,\Born}(\lambda)=\frac{\alpha^2}{s}\left[\sum_i\left(g^i_\lambda\right)^2\left(\frac{s}{u-M_i^2}+\frac{s}{t-M_i^2}\right)\right]^2,\\
\wirk_{\mathrm{\lambda,-\lambda}}&=&\sum_{a=2,3}\wirk_{a,\Born}(\lambda)=\frac{\alpha^2}{s}\left[\left(\sum_i g^i_-g^i_+\frac{t}{u-M_i^2}\right)^2+\left(\sum_i g^i_-g^i_+\frac{u}{t-M_i^2}\right)^2\right].\nn
\label{borncs1}
\end{eqnarray}
Notice that for equal polarizations of the incoming electrons (RR or
LL) the $u$- and $t$-channel matrix elements interfere, whereas in the
other case (RL or LR) no such interference exists.

As a consequence of rotational invariance, the RL and LR
cross sections are equal. Furthermore, the
polarized cross sections (\ref{borncs1}) are symmetric with respect to
$\vartheta=90^\circ$ owing to presence of two identical fermions in
the final state. These symmetries hold to all orders of perturbation
theory. 

For $|t|,|u|\gg\MZ^2$ the polarized cross sections take the simple
form
\beqar\label{approx}
\wirk_{\RR}&\sim&
\frac{\al^2}{s}\frac{16}{\cw^4}\frac{1}{\sin^4\vartheta},\nl
\wirk_{\LL}&\sim& \frac{1}{16\sw^4}\wirk_{\RR},\nl
\wirk_{\LR}&\sim& \frac{(1-\cos\vartheta)^4 + (1+\cos\vartheta)^4}{64}\wirk_{\RR}.
\eeqar 

The integrated cross section is given by
\beq
\si = \frac{1}{2}\int\rd\Omega\wirk =
\int_{\cos\vartheta>0}\rd\Omega\wirk,
\eeq
with the symmetry factor $1/2$ owing to the two identical electrons in the
final state.

\subsection{Polarization asymmetries}

\newcommand{\NRR}{N_{\RR}}
\newcommand{\NRL}{N_{\RL}}
\newcommand{\NLR}{N_{\LR}}
\newcommand{\NLL}{N_{\LL}}
\newcommand{\siRR}{\si_{\RR}}
\newcommand{\siRL}{\si_{\RL}}
\newcommand{\siLR}{\si_{\LR}}
\newcommand{\siLL}{\si_{\LL}}
With polarized electron beams and the possibility to invert these
polarization, four different polarized cross sections
can be measured. Assuming the same luminosity $\L$ for all channels, 
the numbers of events for these polarization settings
are given by:
\beqar
N_{\RR} &=& \L \int \rd\Omega
\left(\frac{\rd\si}{\rd\Omega}\right)(+P_1,+P_2),\nl 
N_{\RL} &=& \L \int \rd\Omega
\left(\frac{\rd\si}{\rd\Omega}\right)(+P_1,-P_2),\nl 
N_{\LR} &=& \L \int \rd\Omega
\left(\frac{\rd\si}{\rd\Omega}\right)(-P_1,+P_2),\nl 
N_{\LL} &=& \L \int \rd\Omega
\left(\frac{\rd\si}{\rd\Omega}\right)(-P_1,-P_2),
\eeqar
where 
the integral extends over a suitable volume of phase space.

{}From these four event rates, three independent asymmetries can be
formed \cite{Cuypers,Cz98}. A particularly suitable set consists of
\beqar\label{asymmexp}
\frac{\NLL+\NLR-\NRL-\NRR}{\NLL+\NLR+\NRL+\NRR} = P_1 A^{(1)}_{\LR}\nlc[1.2ex]
\frac{\NLL+\NRL-\NLR-\NRR}{\NLL+\NLR+\NRL+\NRR} = P_2 A^{(1)}_{\LR}\nlc[1.2ex]
\sqrt{\frac{(\NLL-\NRR)^2-(\NRL-\NLR)^2}{(\NLL+\NRR)^2-(\NLR+\NRL)^2}} = A^{(2)}_{\LR}.
\eeqar
The asymmetry functions 
\beq
A^{(1)}_{\LR}= \frac{\siLL-\siRR}{\siLL+2\siLR+\siRR}
\eeq
and 
\beq
A^{(2)}_{\LR} =
\frac{|\siLL-\siRR|}{\sqrt{(\siLL+\siRR)^2-(2\siLR)^2}}
\eeq
are independent of the polarization degrees. Moreover, for scattering
angles near to $\vartheta=90^\circ$ where $\siLR$ is
suppressed with respect to $\siLL$ and $\siRR$, $A^{(1)}_{\LR}$ and
$A^{(2)}_{\LR}$ are roughly equal to
\beq
A^{(3)}_{\LR} = \frac{\siLL-\siRR}{\siLL+\siRR}.
\eeq
All asymmetry functions $A^{(k)}_{\LR}$, $k=1,2,3$, are proportional to 
\beq
\siLL-\siRR \propto 
\left(g^{Z}_{-}\right)^2 - \left(g^{Z}_{+}\right)^2
\propto 1-4\sw^2,
\eeq
and, owing to $\sw^2\approx0.23\sim1/4$, they are very sensitive to
small variations of the weak mixing angle.  

The set of asymmetries defined in \refeq{asymmexp} is particularly
useful in order to determine the polarization degrees of the two
electron beams and the weak mixing angle. In the third asymmetry the
polarization degrees drop out, and the weak mixing angle can be
directly determined. Then $A^{(1)}_{\LR}$ is fixed and the first two
asymmetries can be used to measure directly the polarization degrees.

\section{Electromagnetic radiative corrections}
\label{se:emcorr}

Using crossing symmetry, the one-loop radiative corrections to \Moeller\ 
scattering can be directly obtained from the results for Bhabha
scattering listed in \citere{Denner1}. The corresponding Feynman
diagrams result 
from those of \citere{Denner1} via
crossing. The results of \citere{Denner1} have been
obtained within the on-shell renormalization scheme \cite{Bohm1} with
the fine-structure constant $\alpha$ and the masses $\MW$, $\MZ$,
$\MH$ and $\Mf$ as physical parameters. 

 In the ultra-relativistic limit 
 the amplitudes for
the virtual corrections factorize into Born matrix elements and
correction factors ($r=u,t$; $i=\ga,Z$; $a=1,2,3$)
\begin{equation}
\mathcal{M}^{ri}_{a,\virt}(\lambda)=
\mathcal{M}^{ri}_a(\lambda) \delta^{ri}_{a,\virt}(\lambda).
\label{matfakt}
\end{equation}
The virtual corrections can be split into electromagnetic and weak
corrections
\begin{equation}
\delta^{ri}_{a,\virt}(\lambda)=
\delta^{ri}_{a,\elect,\virt}(\lambda)+\delta^{ri}_{a,\w,\virt}(\lambda).
\end{equation}
Note that we define this splitting differently from \citere{Denner1}.
Into the electromagnetic corrections we include only those diagrams
(and the corresponding counterterms) that arise from the Born matrix
elements by adding a photon line; all other corrections are considered
as weak corrections. In contrast to \citere{Denner1} the fermionic
loops contained in the photon self-energy are included in the weak
corrections.

\subsection{Electromagnetic virtual corrections}
The electromagnetic virtual corrections consist of vertex (V) and
box diagrams (B) with virtual photons
\begin{equation}
\delta^{ri}_{a,\elect,\virt}(\lambda)=\delta^{ri}_{a,\elect,\Ver}(\lambda)+\delta^{ri}_{a,\elect,\Boxen}(\lambda).
\label{emvirtkorr}
\end{equation}
The corresponding amplitudes are infrared (IR)-divergent. The
IR singularities are regularized by an infinitesimal 
photon mass $\mu$.

For each vertex in the Born diagrams there is a 
vertex-correction diagram. The corresponding correction factors read 
\begin{equation}
\delta^{ri}_{a,\elect,\Ver}(\lambda)=2F_{\elect}(r),
\end{equation}
with the electromagnetic form factor $F_{\elect}$ given in
\citere{Denner1}. Each Born diagram with photon-exchange gives rise to
two box diagrams with double photon exchange and correction factors
\begin{equation}
\delta^{u\ga}_{{1\atop
    2},\elect,\Boxen}(\lambda)=C^{\gamma\gamma}_\pm (u,t),\qquad 
\delta^{t\ga}_{{1\atop
    3},\elect,\Boxen}(\lambda)=C^{\gamma\gamma}_\pm (t,u), 
\end{equation}
and for each Born diagram with $\PZ$-boson exchange four box diagrams 
with $\PZ$-boson and photon exchange and correction factors exist
\begin{equation}
\delta^{uZ}_{{1\atop 2},\elect,\Boxen}(\lambda)=C^{\gamma Z}_\pm
(u,t),\qquad 
\delta^{tZ}_{{1\atop 3},\elect,\Boxen}(\lambda)=C^{\gamma Z}_\pm (t,u).
\end{equation}
The analytical expressions for the electromagnetic box functions
$C^{\gamma\gamma}_\pm (s,t)$ and $C^{\gamma Z}_\pm (s,t)$ can be found
in \citere{Denner1}. To make correct use of these results, recall that
the Mandelstam variables are related by $s+t+u=0$ in the
ultra-relativistic limit.

\subsection{Soft-photon bremsstrahlung} 
\label{chap:brems}

In order to obtain an IR-finite cross section one has to add real
bremsstrahlung. In the soft-photon limit, \ie restricting the energy
of the emitted photon by  $k^0\le\De E\ll \sqrt{s}$, the corresponding
cross section factorizes into the lowest-order cross section and a 
correction factor 
\begin{equation}
\wirk_{a,\Brems}(\lambda)= \delta_{\Brems}\,\wirk_{a,\Born}(\lambda)
\end{equation}
with
\begin{eqnarray}
\delta_\Brems 
&=&\frac{\alpha}{\pi}\left\{4\ln{\frac{2\Delta E}{\mu}}
\left[\ln\frac{ut}{s\Me^2}-1\right] - \left[\ln\frac{s}{\Me^2}-1\right]^2+1
-\frac{2\pi^2}{3}+X\right\}, 
\label{deltabrems} 
\end{eqnarray}
and
\begin{equation}
X=\left(\ln{\frac{u}{t}}\right)^2+\frac{\pi^2}{3}. 
\label{beta}
\end{equation}

Note that, owing to the condition $k_0\leq \Delta E$, this result is
frame-dependent and cannot be directly obtained from the corresponding
result for Bhabha scattering via crossing symmetry.

\subsection{Corrections to the polarized cross sections}
The virtual $\Oa$ corrections to the polarized cross sections 
\refeq{borncs0} are
given by the interference terms between
lowest-order and one-loop matrix elements 
\begin{equation}
\Delta
\wirk_{a,\elect,\virt}(\lambda)=\frac{\alpha^2}{4s}
\sum_{r,i}\sum_{r',i'}\Re\left\{\mathcal{M}^{ri}_a(\lambda)
(\mathcal{M}^{r'i'}_a(\lambda))^*
\left[\delta^{ri}_{a,\elect,\virt}(\lambda)
+(\delta^{r'i'}_{a,\elect,\virt}(\lambda))^*\right]\right\}.
\end{equation}
After adding the soft-photon bremsstrahlung cross section, the
total IR-finite $O(\alpha)$ electromagnetic corrections can be written
as follows:
\begin{eqnarray}
\Delta \wirk_{a,\elect}(\lambda)&=&
\Delta \wirk_{a,\elect,\virt}(\lambda)+\wirk_{a,\Brems}(\lambda)\\
&=&\frac{\alpha^2}{4s}\sum_{r,i}\sum_{r',i'}
\Re\left\{\mathcal{M}^{ri}_a(\lambda)(\mathcal{M}^{r'i'}_a(\lambda))^*
\left[\delta^{ri}_{a,\elect}(\lambda)
+(\delta^{r'i'}_{a,\elect}(\lambda))^*+\gamma\right]\right\},\nn
\end{eqnarray}
with%
\footnote{Note that in the corresponding formula (3.24) in
  \citere{Denner1} the functions $I_5^{\gamma \gamma}(u,t)$ and
  $I_5^{\gamma Z}(u,t)$ should not be multiplied with $\alpha/2\pi$.}
\begin{eqnarray}
\delta^{u\ga}_{{1\atop 2},\elect}(\lambda)&=&
\frac{\alpha}{2\pi}\left[Z+Y(u)+X\right]
\pm 2I_5^{\gamma \gamma}\left(u,{t \atop s}\right),\nonumber\\
\delta^{t\ga}_{{1\atop 3},\elect}(\lambda)&=&
\frac{\alpha}{2\pi}\left[Z+Y(t)+X\right]
\pm 2I_5^{\gamma \gamma}\left(t,{u \atop s}\right),\nonumber\\
\delta^{uZ}_{{1\atop 2},\elect}(\lambda)&=&
\frac{\alpha}{2\pi}\left[Z+Y(u)+X+2D(u,t)\right]
\pm 4I_5^{\gamma Z}\left(u,{t \atop s}\right),\nonumber\\
\delta^{tZ}_{{1\atop 3},\elect}(\lambda)&=&
\frac{\alpha}{2\pi}\left[Z+Y(t)+X+2D(t,u)\right]
\pm 4I_5^{\gamma Z}\left(t,{u \atop s}\right),
\end{eqnarray}
and the cut-off-dependent factor 
\begin{equation}
\gamma=4\frac{\alpha}{\pi}\ln{\frac{2\Delta E}{\sqrt{s}}}
\left[\ln\frac{ut}{s\Me^2}-1\right].
\end{equation}
The functions $I_5^{\gamma \gamma}(u,t)$ and $I_5^{\gamma Z}(u,t)$ are
defined in \citere{Bohm1}, and  the function $D(u,t)$
reads\footnote{In the definition of this function in (3.25)
  in \citere{Denner1} the factor 2 in front of the Spence function
  should be removed.} 
\begin{equation}
D(u,t)=-\ln{\frac{-t}{s}}
\left[\ln{\frac{\MZ^2-u}{-u}}+\ln{\frac{\MZ^2-u}{\MZ^2}}\right] 
+\Sp \left(\frac{\MZ^2+t}{t}\right)- \Sp\left(\frac{\MZ^2+s}{s}\right)
\end{equation}
with the Spence function $\Sp(x)=-\int_0^1
(\mathrm{d}t/t)\ln{(1-xt)}$.
Furthermore, we have introduced
\begin{eqnarray}
Z&=&3\ln{\frac{s}{\Me^2}}+\frac{2\pi^2}{3}-4, \nl
Y(r)&=&-\left(\ln{\frac{-r}{s}} \right)^2-2\ln{\frac{-r}{s}}\ln{\frac{r+s}{s}}+3\ln{\frac{-r}{s}}-\pi^2,
\end{eqnarray}
which correspond to the quantities defined in \citere{Denner1}.
\section{Weak radiative corrections}
\label{se:weakcorr}
\subsection{Corrections to the matrix elements}
The weak radiative corrections result from self-energy ($\Sigma$),
vertex (V) and box (B) diagrams 
\begin{equation}
\delta^{ri}_{a,\w,\virt}(\lambda)=\delta^{ri}_{a,\w,\Sigma}(\lambda)+\delta^{ri}_{a,\w,\Ver}(\lambda)+\delta^{ri}_{a,\w,\Boxen}(\lambda).
\end{equation}

In the on-shell renormalization scheme the electron wave function is
not renormalized, and the only self-energy corrections to \Moeller\ 
scattering come from the exchanged gauge bosons. Each photon exchange
Born diagram gives rise to a photon-self-energy correction term,
and each $\PZ$-boson-exchange Born diagram to three correction terms that
involve the $\PZ$-self-energy, and the $\gamma$--$\PZ$ and
$\PZ$--$\gamma$ mixing energy. The corresponding correction factors
are given by
\begin{eqnarray}
\delta^{r\ga}_{a,\w,\Sigma}(\lambda)&=&\Pi^\gamma(r), \nonumber\\
\delta^{rZ}_{1,\w,\Sigma}(\lambda)&=&\Pi^Z(r)+\frac{2}{g_\lambda ^Z}\,\Pi^{\gamma Z}(r), \nonumber\\
\delta^{rZ}_{{2\atop 3},\w,\Sigma}(\lambda)&=&\Pi^Z(r)+\left(\frac{1}{g_+ ^Z}+\frac{1}{g_- ^Z}\right)\Pi^{\gamma Z}(r),
\end{eqnarray}
with
\begin{equation}
\Pi^\gamma(r)=-\frac{\Sigma^\gamma (r)}{r},\qquad \Pi^Z(r)=-\frac{\Sigma^Z (r)}{r-\MZ^2},\qquad \Pi^{\gamma Z}(r)=-\frac{\Sigma^{\gamma Z}(r)}{r}.
\end{equation}
The self-energies and mixing energies can be found in
\citere{Denner2}. The hadronic part of the photon vacuum polarization
associated with light quarks cannot be calculated reliably within
perturbation theory. Instead, the contribution of the five light
quarks can be directly obtained from the cross section for
$\Pep\Pem\to\mathrm{hadrons}$ via a dispersion integral \cite{Ei95}.  
We use the parameterization
\begin{equation}
\Re \Pi_{\mathrm{had}}^\gamma (s)=A+B\ln{(1+C|s|)},
\end{equation}
with $A,B,C$ adjusted to a recent fit \cite{Burkhardt} of the
experimental data 
(Note that the constants $A,B,C$ are fixed differently in different
regions).

The weak vertex corrections involve, besides contributions from the
exchange of virtual $\PZ$ and \PW~bosons, a non-abelian contribution
and lead to the correction factors
\begin{eqnarray}
\delta^{ri}_{1,\w,\Ver}(\lambda)&=&2F_\w^i(r,\lambda), \nonumber\\
\delta^{ri}_{{2\atop 3},\w,\Ver}(\lambda)&=&F_\w^i(r,+)+F_\w^i(r,-),
\end{eqnarray}
with the weak form factors $F_\w^i(r,\lambda)$ \cite{Denner1}. The
weak box corrections consist of four diagrams with double
$\PZ$-boson exchange and two diagrams with double $\PW$-boson exchange. By
convention they are treated as corrections to the photon-exchange
Born diagrams. The corresponding correction factors read
\begin{eqnarray}
\delta^{u\ga}_{1,\w,\Boxen}(\lambda)&=&(g_\lambda^Z)^4C_+^{ZZ}(u,t)+(g_\lambda^W)^4C_+^{WW}(u,t), \nonumber\\
\delta^{t\ga}_{1,\w,\Boxen}(\lambda)&=&(g_\lambda^Z)^4C_+^{ZZ}(t,u)+(g_\lambda^W)^4C_+^{WW}(t,u), \nonumber\\
\delta^{u\ga}_{2,\w,\Boxen}(\lambda)&=&(g_+^Z)^2(g_-^Z)^2C_-^{ZZ}(u,t),\nonumber\\
\delta^{t\ga}_{3,\w,\Boxen}(\lambda)&=&(g_+^Z)^2(g_-^Z)^2C_-^{ZZ}(t,u),
\end{eqnarray}
while $\delta^{rZ}_{a,\w,\Boxen}(\lambda)=0$. The weak box functions
$C^{ZZ}_\pm (s,t)$ and $C^{WW}_+(s,t)$ are given in 
\citeres{Denner1} and \cite{Bohm1}. 

\subsection{Corrections to the polarized cross sections}
The  weak $\Oa$ corrections to the polarized cross sections
\refeq{borncs0} read 
\begin{equation}
\Delta \wirk_{a,\w}(\lambda)=
\frac{\alpha^2}{4s}\sum_{r,i}\sum_{r',i '}\Re\left\{
\mathcal{M}^{ri}_a(\lambda)(\mathcal{M}^{r'i'}_a(\lambda))^*
\left[\delta^{ri}_{a,\w,\virt}(\lambda)+(\delta^{r'i'}_{a,\w,\virt}(\lambda))^*\right]\right\}.
\label{onelkorr1}
\end{equation}
Finally, the total $\Oa$ corrections to the cross section 
are given by
\begin{equation}
\Delta \wirk_a(\lambda)=\Delta \wirk_{a,\w}(\lambda)+\Delta \wirk_{a,\elect}(\lambda).
\end{equation}

\section{Numeric results and discussion}
\label{se:numeric}

In the previous chapters we have summarized the analytical formulas
for the polarized \Moeller\ cross section including the complete $O(\alpha)$
radiative corrections in the soft-photon approximation. Here, we
present a numerical evaluation of these results obtained with the
masses
\begin{equation}
\MZ=91.187\GeV,\qquad \MW=80.4\GeV,\qquad \MH=300\GeV,
\qquad \Mt=175\GeV,
\end{equation}
and with the fine-structure constant $\alpha=1/137.03604$.

Our main 
interest lies in the discussion of 
the relative weak
corrections defined by $\delta_\w=\Delta\wirk_\w/\wirk_\Born$. These
are sensitive to the complete structure of
the Electroweak Standard Model and may influence the determination of
the weak mixing angle from asymmetries considerably. 
On the other hand, the electromagnetic corrections 
involve only well-known physics but
are sensitive to the experimental cuts for hard-photon emission.
They must be included properly for specific
experiments, but as far as the asymmetries are concerned, they
are relatively  unimportant.  The
soft-photon effects factorize and cancel exactly, and the same should
be true for the bulk of the hard-photon effects. The remaining
electromagnetic effects are mostly
proportional to the lowest-order asymmetries
and therefore do not give rise to large corrections.
We note that the complete electromagnetic corrections for the luminosity
measurement have been discussed in \citere{qedmc}.

The weak corrections are separated in a 
gauge-invariant way into fermionic and bosonic parts
\begin{equation}
\delta_{\w} =\delta_{\w,\fer}+\delta_{\w,\bos}.
\end{equation}
The fermionic part consists of all diagrams with closed 
fermion loops and
fermionic counterterms and is, in the renormalization scheme of
\citere{Bohm1},
given by the hadronic and leptonic contributions to the gauge-boson
self-energies. The remaining weak
corrections are called bosonic and are, within the 't Hooft--Feynman
gauge, further split into self-energy, vertex, and box contributions,
\begin{equation}
\delta_{\w,\bos} =
\delta_{\w,\bos,\Sigma}+\delta_{\w,\bos,\Ver}+\delta_{\w,\bos,\Boxen}.
\end{equation}
The fermionic corrections involve, in particular, the contributions that
are related to the
running of the electromagnetic coupling constant $\al$. Including only
the contributions of the light fermions, we define the corresponding
correction factors to the matrix elements $\M^{ri}_a(\la)$ as 
\begin{equation}\label{runcorr}
\delta^{r\ga}_{a,\run}(\lambda)=\Pi_{f\ne\mathrm{top}}^\gamma(r), \qquad
\delta^{rZ}_{a,\run}(\lambda)=\Pi_{f\ne\mathrm{top}}^\gamma(\MZ^2), \qquad r=u,t. 
\end{equation}
Then, the fermionic corrections can be split into those related to the
running of $\al$ and the remaining ones
\begin{equation}
\delta_{\w,\fer} =\delta_{\w,\run}+\delta_{\w,\fer,\rem}.
\end{equation}

\begin{figure}
\centerline{
\setlength{\unitlength}{1cm}
\begin{picture}(10,7.8)
\put(0,0){\includegraphics{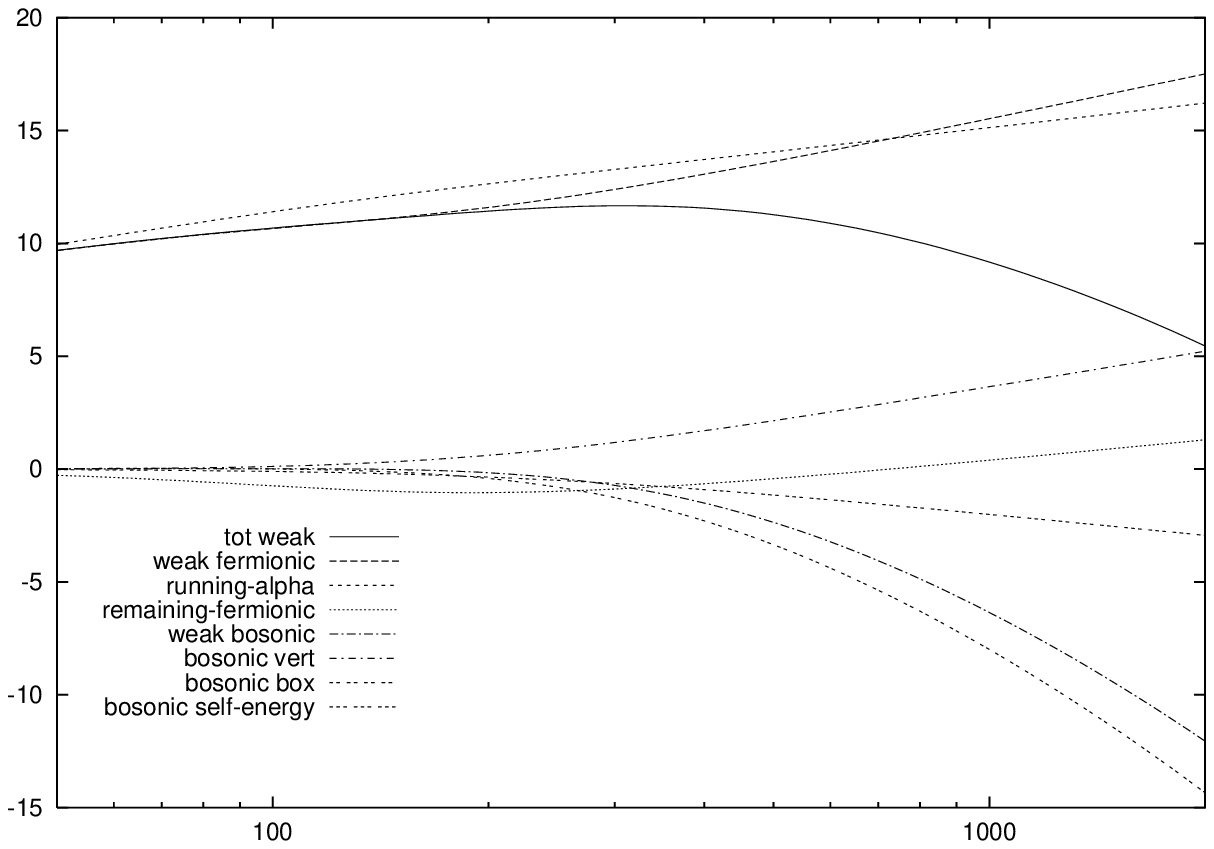}}
\put(2,-0.1){\makebox(6,0.5)[b]{$\sqrt{s}\,[\GeV]$}}
\put(-3,4){\makebox(1.5,1)[r]{$\delta_\w\,[\%]$}}
\end{picture}}
\caption[feb8]{Individual contributions to the relative weak
  corrections to the unpolarized differential cross section for
  $\vartheta=90^\circ$}
\label{plot5}
\end{figure}

The different contributions to the weak corrections for the
unpolarized cross section $(P_1=P_2=0)$ are presented in \reffi{plot5} for a 
scattering angle of $90^\circ$ and CM energies between $50\GeV$  
and $2\TeV$.
The complete weak corrections are of the order 10\% and vary slowly with 
energy (in the range between $+5.4$\% and $+11.7$\%). For energies
below the $\PZ$-boson mass they are dominated by the fermionic
contributions, which grow logarithmically from 9.7\% at $50\,\GeV$ to
17.5\% at $2\TeV$. 
The bulk of the fermionic corrections is due to the
running of $\al$ (ranging between 10.0\% and 16.2\%), and the
remaining fermionic corrections are less than 1.3\% in the considered
energy range.
While the bosonic corrections are less than 1\%
below $300\GeV$, they reduce the weak corrections considerably 
at higher energies, 
\eg by 12\% at $2\TeV$. This reduction originates
predominantly from the box corrections, more precisely from the boxes
involving \PW-boson exchange.  The weak vertex corrections yield a
positive contribution (5.2\% at 2 TeV) which is again dominated by the
diagrams involving  \PW~bosons. 
 Finally, the bosonic self-energy
corrections contribute $-2.9$\% at $2\TeV$.

In order to allow for a detailed check of our calculations we
present in \refta{table1} the differential and integrated
lowest-order cross sections together with the relative
electromagnetic and weak corrections split into fermionic and bosonic
parts. The fermionic corrections are further divided into those
originating from the running of $\al$ and the rest.
\begin{table}
\newdimen\digitwidth
\setbox0=\hbox{0}
\digitwidth=\wd0
\catcode`!=\active
\def!{\kern\digitwidth}
\newdimen\minuswidth
\setbox0=\hbox{$-$}
\minuswidth=\wd0
\catcode`?=\active
\def?{\kern\minuswidth}
\begin{center}
\arraycolsep 6pt
$$\begin{array}{|c|c||c|c|c|c|c|c|c||}
\hline
\sqrt{s} & \vartheta & \wirk_{\Born\strut} &
\delta_{\elect} &
\delta_{\w} &
\delta_{\w,\bos}  &
\delta_{\w,\fer}  &
\delta_{\w,\run}  &
\delta_{\w,\fer,\rem}  
\\{}
[\mathrm{GeV}] & &[\mathrm{pb}] &
[\%] &[\%] &[\%] &[\%] &[\%] &[\%] \\
\hline\hline
   & 10^\circ &  36001\phantom{.}&  -26.95  &  !6.98  
&  !-0.00 &  !6.98 &  !6.99  & !-0.01  \\
 \cline{2-9}
   & 30^\circ &  476.75  &  -30.22  &   !9.25  
&  !-0.01 &  !9.25 &  !9.36 &  !-0.11  \\
\cline{2-9}
 !100 & 90^\circ &  23.024  &  -32.82  &   10.68  
& ?!0.02  &  10.66 &  11.40 &  !-0.74  \\
\cline{2-9}
      & 10^\circ< \vartheta <90^\circ &   3461.1      &  -28.42 &  ! 7.98
&  !-0.00 &  !7.99 &  !8.05 &  !-0.06  \\
\cline{2-9}
      & 30^\circ< \vartheta <90^\circ &  402.08      &  -31.39  &  !9.96  
&  !-0.00 &  !9.96 &  10.27 &  !-0.32  \\
\hline\hline
      & 10^\circ &   1452.9   &  -31.59      &   10.03      
&  !-0.19 &  10.22 &  10.36 &  !-0.14 \\
\cline{2-9}
      & 30^\circ & 21.326  &  -34.85 &   10.89 
&  !-1.16 &  12.05 &  12.67 &  !-0.62 \\
\cline{2-9}
 !500 & 90^\circ & 1.2366  &  -37.31 &   11.27 
&  !-2.36 &  13.63 &  14.06 &  !-0.43 \\
\cline{2-9}
      & 10^\circ< \vartheta <90^\circ &  145.61 &  -33.13  &  10.53  
&  !-0.60 &  11.13 &  11.45 &  !-0.33 \\
\cline{2-9}
 & 30^\circ< \vartheta <90^\circ &  19.533  &  -36.02  &  11.00  
&  !-1.76 &  12.77 &  13.37 &  !-0.60 \\
\hline\hline
      & 10^\circ &  98.713  &  -35.58  &   !8.27  
&  !-4.29 &  12.56 &  13.15 &  !-0.59  \\
\cline{2-9}
      & 30^\circ &  1.4375  &  -38.78  &   !5.11  
&  -10.04 &  15.16 &  15.11 &  ?!0.05  \\
\cline{2-9}
 2000 & 90^\circ &  !0.07957  &  -41.07  &   !5.46  
&  -12.05 &  17.51 &  16.21 &  !?1.30 \\
\cline{2-9}
      & 10^\circ< \vartheta <90^\circ &  9.9259  &  -37.08  &   !6.80 
&  !-6.95 &  13.74 &  14.07 &  !-0.33  \\
\cline{2-9}
      & 30^\circ< \vartheta <90^\circ &  1.2867  &  -39.88  &  !5.02  
&  -11.23 &  16.25 &  15.68 &  ?!0.57  \\
\hline
\end{array}$$
\caption{\label{table1}Lowest-order cross sections and relative 
corrections for unpolarized particles} 
\label{ta:born}
\end{center}
\end{table}%
We recall that the
electromagnetic corrections are evaluated in the soft-photon
approximation and depend strongly 
on the soft-photon-energy cut-off
which was set to $\Delta E =0.05 \sqrt{s}$. They are
strongly reduced once hard bremsstrahlung is included.

\begin{figure}
\centerline{
\setlength{\unitlength}{1cm}
\begin{picture}(10,7.8)
\put(0,0){\includegraphics{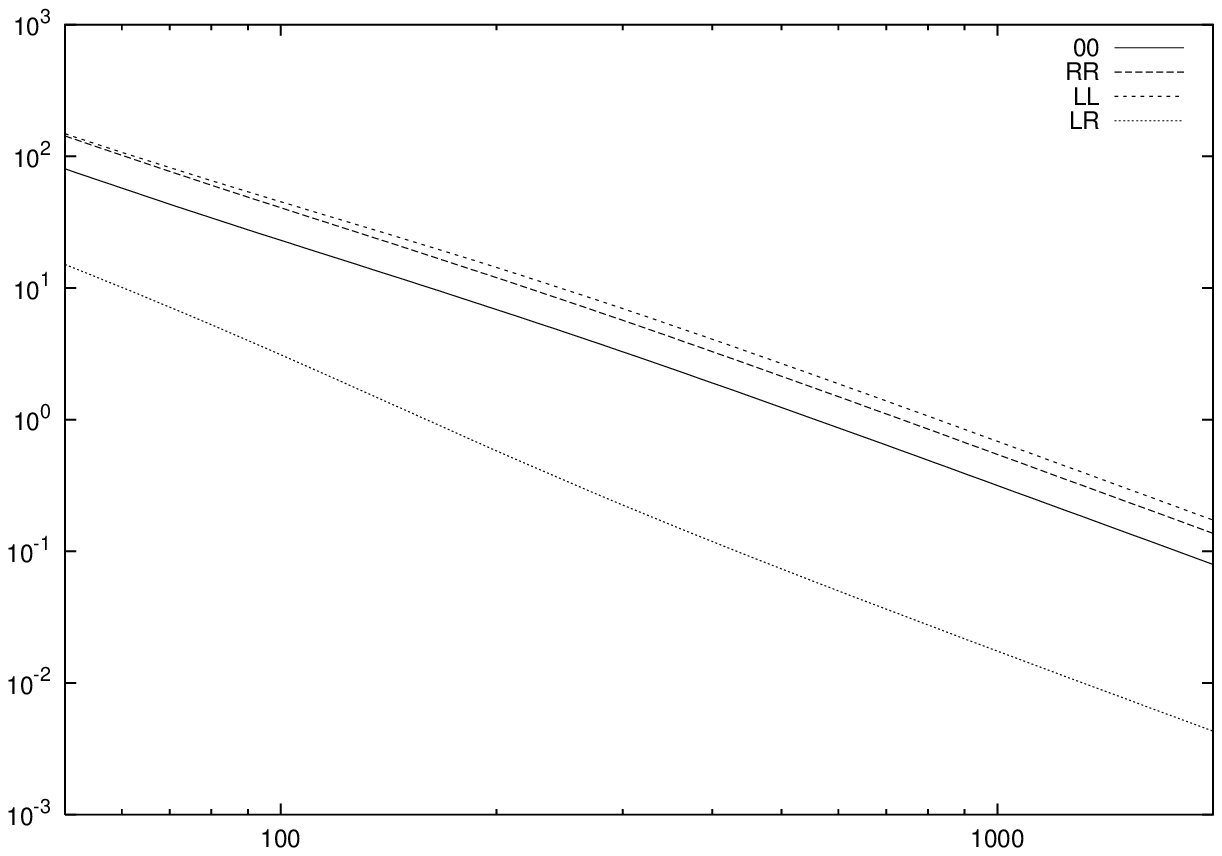}}
\put(2,-0.1){\makebox(6,0.5)[b]{$\sqrt{s}\,[\GeV]$}}
\put(-3,4){\makebox(1.5,1)[l]{$\wirk\,[\mathrm{pb}]$}}
\end{picture}}
\caption{\label{plot3} Unpolarized (00) and polarized (RR,LL,LR)  differential cross sections in Born approximation ($\vartheta=90^\circ$)}
\end{figure}

\begin{figure}
\centerline{
\setlength{\unitlength}{1cm}
\begin{picture}(10,7.8)
\put(0,0){\includegraphics{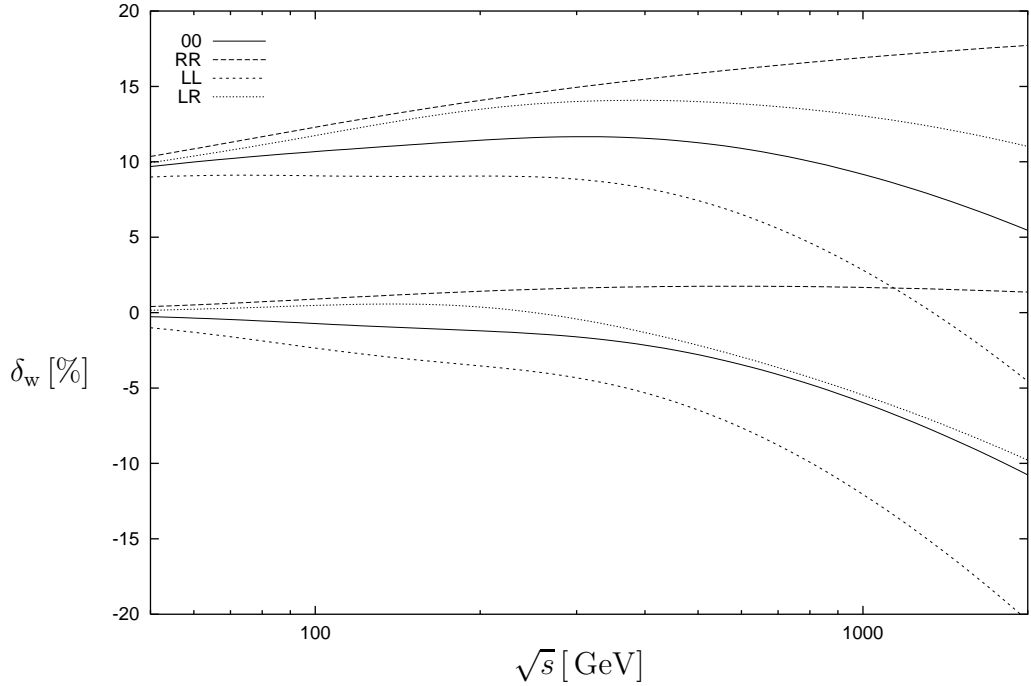}}
\put(2,-0.1){\makebox(6,0.5)[b]{$\sqrt{s}\,[\GeV]$}}
\put(-3,3.5){\makebox(1.5,1)[r]{$\delta_\w \,[\%]$}}
\end{picture}}
\caption{Relative weak corrections to the unpolarized (00) and
  polarized (RR, LL, LR) differential cross sections
  including (upper curves) and excluding (lower curves) the effect of
  the running $\al$  ($\vartheta=90^\circ$)} 
\label{plot4}
\end{figure}

In \reffis{plot3} and \ref{plot4} we show the lowest-order
cross sections and the corresponding relative weak corrections
for various polarizations of the incoming electrons at the
scattering angle $\vartheta=90^\circ$ as a function of the CM
energy.
For $\sqrt{s}\gg\MZ$, the lowest-order cross sections drop as $1/s$, and
the ratios between different polarized cross sections are
energy-independent [\cf\refeq{approx}].  
The relative corrections are about 10\% for small energies and
vary between -5\% and 18\% at $2\TeV$ depending on the polarization.
Besides the complete weak corrections we show in \reffi{plot4} also
the weak corrections that remain after the subtraction of the effects
of the running $\al$ [see \refeq{runcorr}]. These remaining weak
corrections are small at low energies for all polarizations and for
purely right-handed electrons at all energies.  The large negative
corrections for the other polarizations at high energies are due to
vertex and box diagrams involving \PW-boson exchange, which contribute
only for at least one or two left-handed incoming electrons,
respectively.  The small differences in the corrections for different
polarizations at low energies are due to the $\gamma$--$\PZ$
mixing-energy and to differences in the interference structures in
(\ref{borncs1}).  

\begin{figure}
\centerline{
\setlength{\unitlength}{1cm}
\begin{picture}(10,7.8)
\put(0,0){\includegraphics{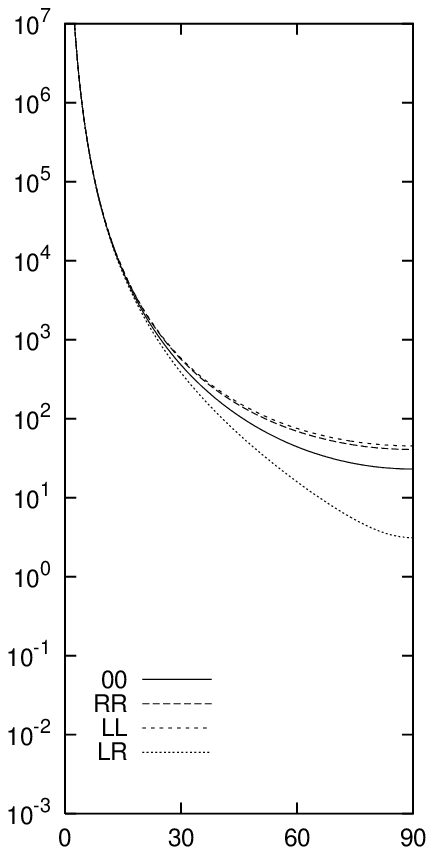}}
\put(4,0){\includegraphics{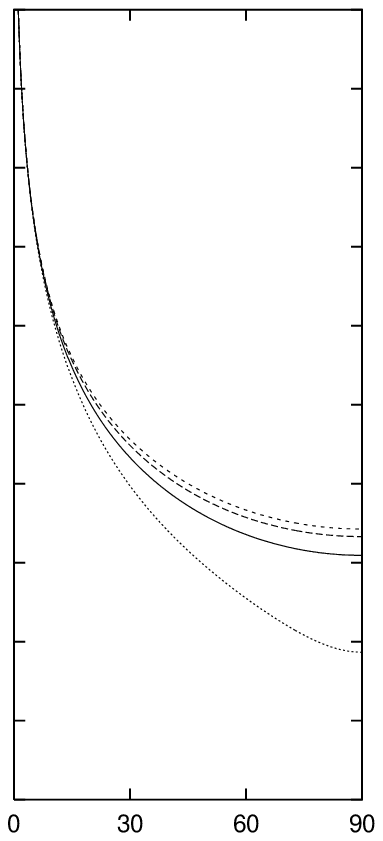}}
\put(8,0){\includegraphics{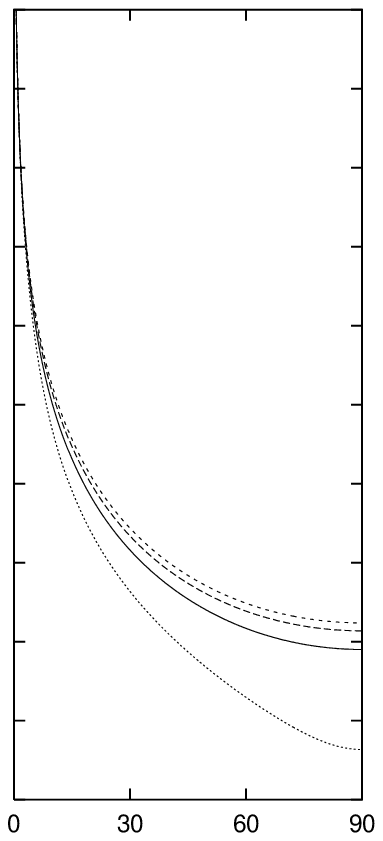}}
\put(-3,4){\makebox(1.5,1)[r]{$\wirk \,[\mathrm{pb}]$}}
\put(-0.7,9){\makebox(3.5,0.5)[t]{$\sqrt{s}=100\,\GeV$}}
\put(3.3,9){\makebox(3.5,0.5)[t]{$\sqrt{s}=500\,\GeV$}}
\put(7.3,9){\makebox(3.5,0.5)[t]{$\sqrt{s}=2000\,\GeV$}}
\put(-0.7,-0.2){\makebox(3.5,0.5)[t]{$\vartheta\, [^\circ]$}}
\put(3.3,-0.2){\makebox(3.5,0.5)[t]{$\vartheta\,[^\circ]$}}
\put(7.3,-0.2){\makebox(3.5,0.5)[t]{$\vartheta\,[^\circ]$}}
\end{picture}}
\caption{\label{plot1}Unpolarized (00) and polarized (RR,LL,LR)
  differential Born cross section as a function of the scattering
  angle $\vartheta$ for different CM energies}
\end{figure}

\begin{figure}
  \centerline{ \setlength{\unitlength}{1cm}
\begin{picture}(10,7.8)
\put(0,0){\includegraphics{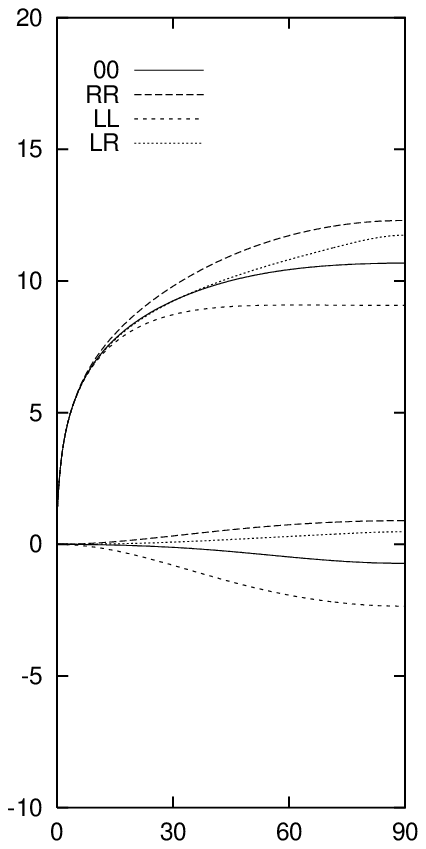}}
\put(4,0){\includegraphics{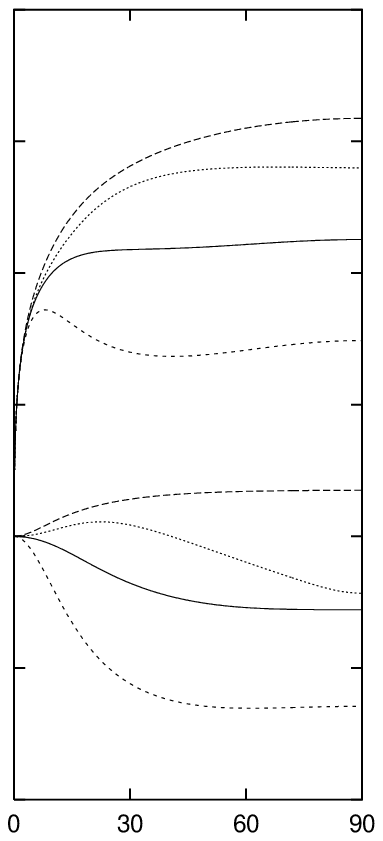}}
\put(8.5,0){\includegraphics{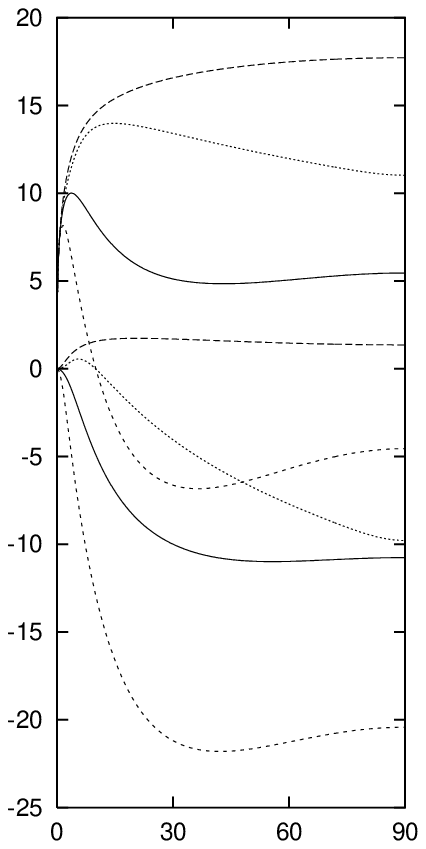}}
\put(-3,4){\makebox(1.5,1)[r]{$\delta_\w \,[\%]$}}
\put(-0.7,9){\makebox(3.5,0.5)[t]{$\sqrt{s}=100\,\GeV$}}
\put(3.3,9){\makebox(3.5,0.5)[t]{$\sqrt{s}=500\,\GeV$}}
\put(7.7,9){\makebox(3.5,0.5)[t]{$\sqrt{s}=2000\,\GeV$}}
\put(-0.7,-0.2){\makebox(3.5,0.5)[t]{$\vartheta\, [^\circ]$}}
\put(3.3,-0.2){\makebox(3.5,0.5)[t]{$\vartheta\,[^\circ]$}}
\put(7.8,-0.2){\makebox(3.5,0.5)[t]{$\vartheta\,[^\circ]$}}
\end{picture}}
\caption{\label{plot2}Relative weak corrections to the unpolarized
  (00) and polarized (RR,LL,LR) differential cross sections as a
  function of the scattering angle $\vartheta$ for different CM
  energies including (upper curves) and excluding (lower
  curves) the effects of the running~$\al$}
\end{figure}

Figures \ref{plot1} and \ref{plot2} illustrate the angular
dependence of the polarized lowest-order cross sections and the
corresponding corrections, respectively. 
Owing to Fermi symmetry the
cross sections are forward--backward symmetric, and it is sufficient to
consider the forward direction.  
In the very forward direction the lowest-order cross section turns into
the Rutherford cross section that diverges for $\theta\to0$ like 
$1/t^2$ and is
independent of the polarization of the incoming electrons. 
As can be seen from \refeq{approx}, for high energies and not too
small scattering angles ($|t|\gg\MZ^2$),
the ratio between the different
polarized lowest-order cross sections takes a particularly simply form. 
Thus, the ratio of LL to RR is constant and roughly
equal to one, and the ratio of LR to RR decreases from 1 at
$\vartheta=0^\circ$ to $1/32$ at $\vartheta=90^\circ$. 
The relative
weak corrections vanish exactly in the forward direction.
This can be explained as follows: In the forward direction, the
lowest-order cross section is determined solely by the $t$-channel
photon-exchange diagram which involves the $1/t$ pole. Similarly, only
corrections that involve this pole are relevant, \ie the self-energy
and vertex corrections to the dominating lowest-order diagram.
However, at the pole the virtual photon in these diagrams becomes real
and these corrections vanish owing to the on-shell renormalization
condition for the electric charge.  
For finite scattering angles, the weak corrections and their
composition behave roughly as for the case of
$90^\circ$ scattering discussed above.

\begin{figure}
\centerline{
\setlength{\unitlength}{1cm}
\begin{picture}(10,7.8)
\put(0,0){\includegraphics{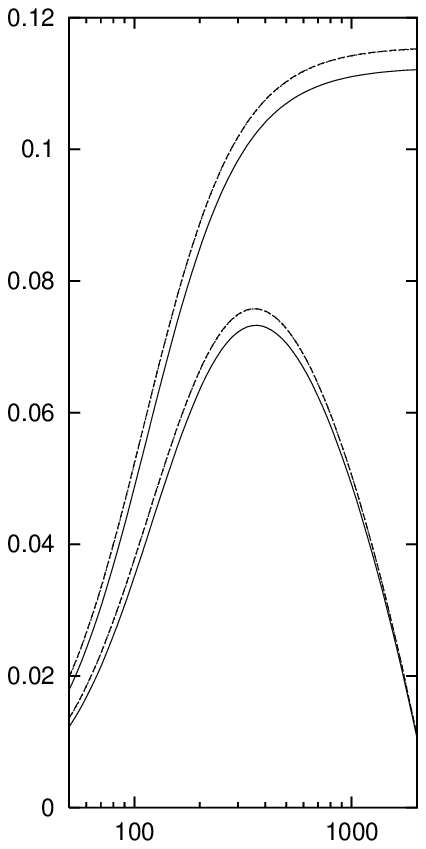}}
\put(4,0){\includegraphics{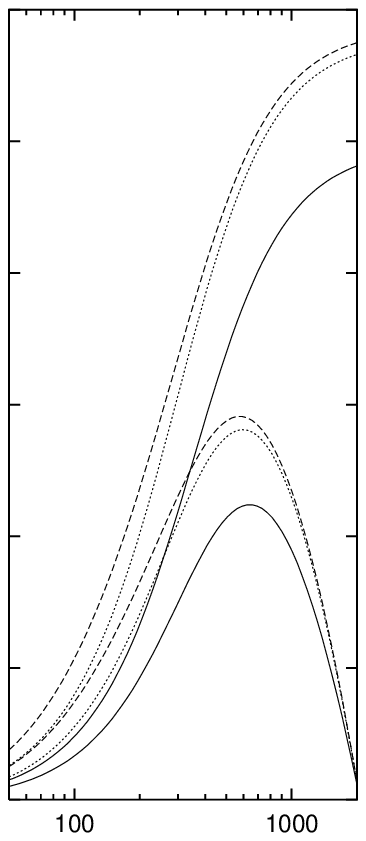}}
\put(8,0){\includegraphics{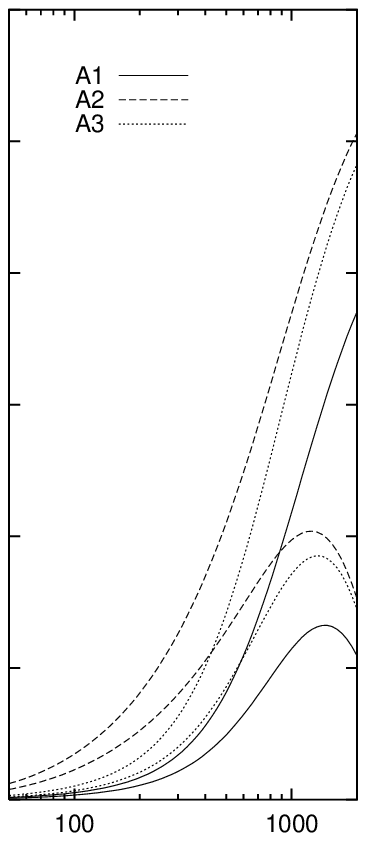}}
\put(-3,4){\makebox(1.5,1)[r]{$A^{(k)}_\LR$}}
\put(-0.7,9){\makebox(3.5,0.5)[t]{$\vartheta =90 ^\circ$}}
\put(3.3,9){\makebox(3.5,0.5)[t]{$\vartheta =30 ^\circ$}}
\put(7.3,9){\makebox(3.5,0.5)[t]{$\vartheta =10 ^\circ$}}
\put(-0.7,-0.2){\makebox(3.5,0.5)[t]{$\sqrt{s}\,[\GeV]$}}
\put(3.3,-0.2){\makebox(3.5,0.5)[t]{$\sqrt{s}\,[\GeV]$}}
\put(7.3,-0.2){\makebox(3.5,0.5)[t]{$\sqrt{s}\,[\GeV]$}}
\end{picture}}
\caption{Differential polarization asymmetries $A^{(k)}_\LR$ in Born
  approximation (upper curves) and including the complete weak 
  one-loop corrections (lower curves) for different scattering angles}
\label{plota3}
\end{figure}

\begin{figure}
\centerline{
\setlength{\unitlength}{1cm}
\begin{picture}(10,7.8)
\put(0,0){\includegraphics{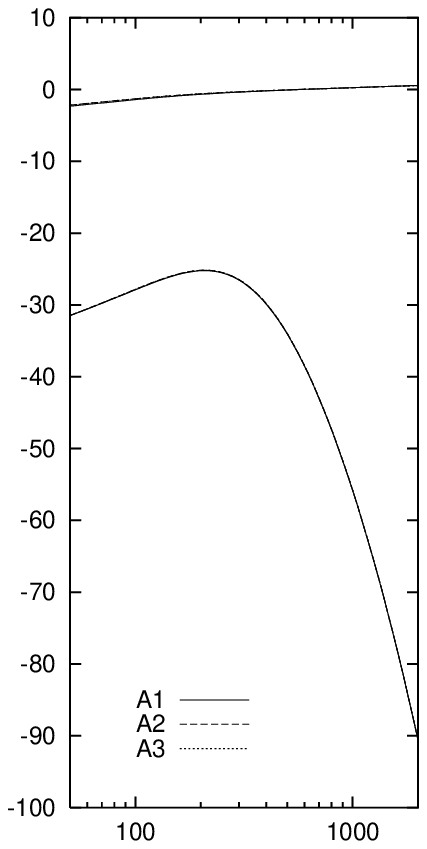}}
\put(4,0){\includegraphics{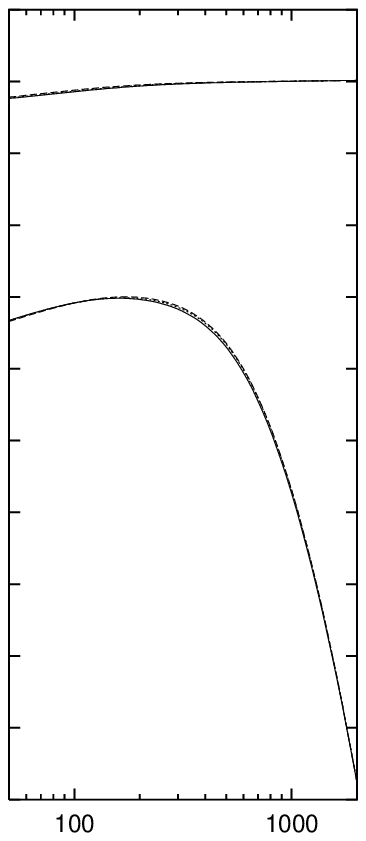}}
\put(8,0){\includegraphics{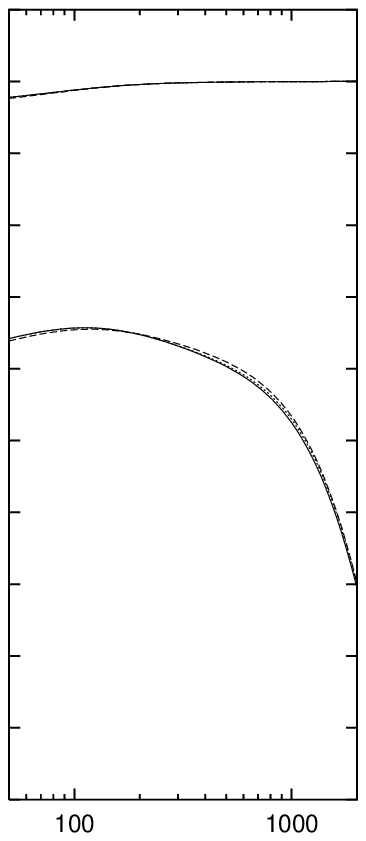}}
\put(-3,5){\makebox(1.5,1)[r]{$\frac{\strut\de A^{(k)}_\LR}{\strut A^{(k)}_\LR}$}}
\put(-3,4){\makebox(1.5,1)[r]{$[\%]$}}
\put(-0.7,9){\makebox(3.5,0.5)[t]{$\vartheta =90 ^\circ$}}
\put(3.3,9){\makebox(3.5,0.5)[t]{$\vartheta =30 ^\circ$}}
\put(7.3,9){\makebox(3.5,0.5)[t]{$\vartheta =10 ^\circ$}}
\put(-0.7,-0.2){\makebox(3.5,0.5)[t]{$\sqrt{s}\,[\GeV]$}}
\put(3.3,-0.2){\makebox(3.5,0.5)[t]{$\sqrt{s}\,[\GeV]$}}
\put(7.3,-0.2){\makebox(3.5,0.5)[t]{$\sqrt{s}\,[\GeV]$}}
\end{picture}}
\caption{\label{plotar}Relative complete weak (lower curves) and
  electromagnetic (upper curves)
  one-loop corrections to the differential polarization asymmetries
  $A^{(k)}_\LR$ for different scattering angles}
\end{figure}

In \reffis{plota3} and \ref{plotar} we show the differential
polarization asymmetries $A^{(k)}_\LR$, $k=1,2,3$,
 in Born approximation and including weak one-loop corrections.
The lowest-order asymmetries are very small for low energies, grow in
the energy range where $|t|\sim\MZ$ and approach constant values
for $|t|\gg \MZ$.  For $A^{(3)}_\LR$ the
asymptotic value is independently of the scattering angle given by
$(1-16\sw^4)/(1+16\sw^4)\approx 0.1156$.  
The three asymmetries have similar energy dependence, and $A^{(2)}_\LR$
is close to $A^{(3)}_\LR$ in particular for $\vartheta=90^\circ$.  The
relative complete weak corrections, shown in \reffi{plotar}, are practically
equal for all three asymmetries (the three curves in \reffi{plotar}
can hardly be distinguished) and are about $-40\%$ for energies up to
a few hundred GeV.
In this regime the asymmetry is mainly due to the $\ga$--$\PZ$ mixing
energy, while the effect of the running $\al$ (which is included in
\reffi{plotar}) cancels to a large extent. 
Owing to the strong
parity-violating effects of the weak bosonic corrections the asymmetries
are even further reduced at higher energies and reverse sign
above about $2\TeV$. 
The large corrections to the asymmetries arise from the fact, that
many of the weak corrections like the $\ga$--$\PZ$ mixing energy or the
contributions involving virtual \PW~bosons are not suppressed by
$1-4\sw^2$ as the lowest-order asymmetries \cite{Cz96}. As a
consequence these corrections are enhanced by a factor of about 10
in the asymmetries.  Some numerical results for the asymmetries
$A^{(k)}_\LR$ in lowest-order approximation and including the weak
corrections are listed in \refta{table2}.  As illustrated in
\reffi{plotar}, the purely electromagnetic corrections affect the
asymmetries only weakly.
\begin{table}
\newdimen\digitwidth
\setbox0=\hbox{0}
\digitwidth=\wd0
\catcode`!=\active
\def!{\kern\digitwidth}
\newdimen\minuswidth
\setbox0=\hbox{$-$}
\minuswidth=\wd0
\catcode`?=\active
\def?{\kern\minuswidth}
\begin{center}
\arraycolsep 6pt
$$\begin{array}{|c|c||c|c|c|c|c|c|}
\hline
\sqrt{s}\,[\mathrm{GeV}] & \vartheta & A^{(1)}_{\LR,\Born} &
A^{(1)}_{\LR,\w} &
A^{(2)}_{\LR,\Born} &
A^{(2)}_{\LR,\w} &
A^{(3)}_{\LR,\Born} &
A^{(3)}_{\LR,\w} \\
\hline\hline
       & 10^\circ &  1.058      &   0.695      &    6.934      &    4.536      &
    2.067      &    1.357      \\
 \cline{2-8}
       & 30^\circ &    9.610      &    6.648      &    21.46      &    14.84
  &    16.01      &    11.07      \\
\cline{2-8}
  !100 & 90^\circ &  48.72      &    35.14      &    52.39      &    37.81
&    52.25      &    37.71      \\
\cline{2-8}
       & 10^\circ< \vartheta <90^\circ &     5.042      &    3.536      &    15
.77      &    11.00      &    9.149      &    6.410      \\
\cline{2-8}
       & 30^\circ< \vartheta <90^\circ & 23.33      &    16.60      &    34.94
     &    24.84      &    32.27      &    22.95      \\
\hline\hline
       & 10^\circ &   16.33      &    9.848      &    42.08      &    25.65
 &    28.39      &    17.16      \\
\cline{2-8}
       & 30^\circ &    67.89      &    42.77      &    90.53      &    57.52
  &    86.91      &    55.08      \\
\cline{2-8}
 !500  & 90^\circ &   107.0      &    70.55      &    110.3      &    72.80
 &    110.2      &    72.77      \\
\cline{2-8}
       & 10^\circ< \vartheta <90^\circ &   38.66      &    24.19      &    66.8
2      &    42.15      &    57.93      &    36.39      \\
\cline{2-8}
       & 30^\circ< \vartheta <90^\circ &  87.16      &    55.93      &    101.6
      &    65.49      &    100.4      &    64.68      \\
\hline\hline
       & 10^\circ &  74.12      &    21.84      &    101.3      &    30.53
&    96.54      &    28.89      \\
\cline{2-8}
       & 30^\circ &  96.26      &    2.246      &    115.0      &    2.730
&    113.2      &    2.679      \\
\cline{2-8}
 2000  & 90^\circ &     112.1      &    10.58      &    115.3      &    10.89
   &    115.2      &    10.89      \\
\cline{2-8}
       & 10^\circ< \vartheta <90^\circ &   86.06      &    14.04      &    109.
1      &    18.16      &    106.1      &    17.57      \\
\cline{2-8}
       & 30^\circ< \vartheta <90^\circ &   102.9      &    3.421      &    115.
1      &    3.866      &    114.4      &    3.837      \\
\hline
\end{array}$$
\caption{Left-right asymmetries in Born approximation $A_{\LR,\Born}$ and
  including weak one-loop corrections $A_{\LR,\w}$ in units of
  $10^{-3}$} 
\label{table2}
\end{center}
\end{table}%

For $s=0.05\GeV^2$ and $\vartheta=90^\circ$ Czarnecki and Marciano
have found corrections to the asymmetry $A^{(3)}_{\LR}$ of $-40\pm3\%$
using the $\overline{\mathrm{MS}}$ scheme, taking the Fermi constant
$\GF$ as input, 
and including electromagnetic corrections. Translating this result to
the on-shell scheme with $\MW$ as input by using the relations given
in \cite{PDG} and subtracting the electromagnetic corrections yields
$-49\%$ whereas we find $-48\%$ in good agreement.

\section{Summary}

We have calculated the $O(\al)$ radiative corrections to the
process $\Pem\Pem \rightarrow\Pem\Pem$ within the Electroweak Standard
Model using the soft-photon approximation for real photon emission.
All relevant analytical results for the corrections to the
polarized cross section have been given.  The corrections have been
split in a gauge-invariant way into electromagnetic and weak ones, and
we have focused on the weak corrections.

The weak corrections to the unpolarized cross section are typically
of the order 10\%. After subtracting the effect of the running of the
electromagnetic coupling constant, the remaining corrections are small at low
energies.
The corrections to the cross section for purely 
right-handed electrons are dominated by the fermion-loop contribution
to the gauge-boson self-energies, in particular by the contributions
associated with the running electromagnetic coupling,
and grow logarithmically with the centre-of-mass energy. For
left-handed electrons the corrections
involve in addition vertex and box diagrams with virtual $\PW$ bosons, which
yield sizeable negative corrections at energies higher than the
$\PZ$-boson mass. 
 
The polarization asymmetry, which ranges between 0 and 0.12 in lowest
order, gets relative corrections of the order $-40\%$ for CM energies
up to $500\GeV$.  For higher energies the parity-violating corrections
increase fast and change the sign of the asymmetry above $2\TeV$.

\end{document}